# Comments on "What the Vector Potential Describes" by E. J. Konopinski

Paul J Cote and Mark A. Johnson

Benet Laboratories
1 Buffington Street
Watervliet, NY, 12189

paul.j.cote@us.army.mil; mark.a.johnson1@us.army.mil

Abstract

The seminal paper on the meaning of the vector potential by E. J. Konopinski is revisited. The significance of this work has not been generally recognized to date. We discuss the underlying assumptions in Konopinski's analysis and show that many of his key results can be obtained from a simpler approach. We then discuss the additional implications of his analysis, which were overlooked by Konopinski himself.

A review of current textbooks and articles on classical electromagnetism indicates that the significance of Konopinski's analysis[1] of the vector potential has been ignored to date. The aim of this note is to alert the scientific community of the full implications of this important paper by a distinguished physicist.

Konopinski's analysis demonstrates the fallacy of the generally held view that the vector potential $A$ has no physical meaning in classical electromagnetism. His paper follows up on Feynman's complaint that a bias exists regarding the vector potential [2]. Konopinski attributes this bias to erroneous notions arising from the gauge concept. By contrast, it is generally accepted from the Bohm-Aharonov [2] effect that $A$ has physical meaning as a real "field" in quantum mechanics (assuming the Coulomb gauge).

The Coulomb potential $\varphi$ has never experienced any such bias about its reality as a field. Its interpretation as the stored potential energy of a unit test charge in an existing field is generally accepted. Invoking the conservation of energy, one can, in principle, experimentally map out a scalar field, at least for the static or quasi-static case. So the scalar potential has physical meaning, and its magnitude at all points in space can be determined experimentally, which satisfies the requirement for treating the scalar potential as a real field.

We first discuss in detail the underlying assumptions in Konopinski's analysis. We consider some elementary issues regarding the choice of gauge that are glossed over in the literature. The aim here is to address questions that might divert attention from the significance of his results. Electromagnetism is a gauge theory, which means a choice of gauge is required to obtain precise, quantifiable, expressions for the relevant variables. (Usually a choice of a zero is the simplest, in analogy with the choice of a zero for the electrostatic potential.) Exercising different gauge choices alters the way in which relationships are expressed, but the physics is unaffected.

The Coulomb potential, $\varphi$, is the standard, operationally defined potential which can be directly measured, or, as in the Lienard-Weichert potentials, computed from first principles in the dynamic case. The magnetic field $B$ is given by,

$$B = \nabla \times A \qquad (1)$$

Thus, the vector potential is only defined to within an arbitrary gradient function, so a gauge choice is required. One source of confusion in the literature is the vector potential is treated as the sole variable requiring gauge choice. There is a second gauge to consider.

The total electric field, (sum of Coulomb and induced fields) is given by,

$$E = E_C + E_I = -\nabla \varphi - \partial A / \partial t \qquad (2)$$

with

$$E_I = -\partial A / \partial t. \qquad (3)$$

The origin of the standard expression in Eq. (3) arises from Maxwell's equation,

$$\nabla \times E_I = -\partial B / \partial t = -\nabla \times \partial A / \partial t, \qquad (4)$$

so that,

$$\nabla \times (E_I + \partial A / \partial t) = 0. \qquad (5)$$

Consequently,

$$E_I + \partial A / \partial t = \nabla \varphi_I, \qquad (6)$$

Again, the curl defines the quantity in parenthesis to within an arbitrary gradient function, $\nabla \varphi_I$, so the standard expression for $E_I$ originates from choosing $\nabla \varphi_I = 0$. Thus, the use of the standard expression in Eq. (3) implies exercising a choice of gauge for $E_I$. Since Eq. (3) represents standard usage in the physics literature, $\nabla \varphi_I = 0$ is the de facto standard gauge in electromagnetism. (The use of Eq.(3) as an expression of the Faraday law has the advantage over the standard line integral formulation in that it is more general and avoids the outmoded action-at-a-distance concept inherent in the line integral form.)

These two choices are not independent. Once a choice for one of the gauges has been exercised, the second "choice" is determined by the laws of physics. To illustrate, if one begins with the standard gauge $\nabla \varphi_I = 0$, then the dynamic form of Gauss' law gives $\nabla \cdot E_I = 0$ in the absence of dynamic Coulomb fields, which, according to Eq.(3), eliminates freedom of choice for the vector potential gauge (We will return to this point later.).

A fact that has escaped notice is that if one begins, instead, by assuming the Coulomb gauge in Eq.(1), as is usually done, then Eq.(6) applies as the correct expression of the Faraday law (with $\nabla \varphi_I$ undetermined), not Eq.(3). The appropriate next step is to invoke Gauss' law ($\nabla \cdot E_I = 0$) which makes it clear that $\nabla \varphi_I = 0$ is required by the laws of physics. Note that in the absence of Coulomb fields, both of the choices considered here give the result that Eq.(3) is a law of physics in their two respective gauges. So the common practice of invoking the Coulomb gauge and the standard gauge, implicit in Eq.(3), does not cause a problem. In the general case, these same two choices give dramatically different results, however. This is the source of error in Jackson's textbook proof [4] that dynamic Coulomb fields propagate faster than the speed of light.

Employing Eq.(3) as a law of physics, Konopinski showed that the vector potential, in the solenoid case, has both physical meaning and measurability. He demonstrated that the generalized momentum, $p = Mv + qA$, is a conserved quantity along any path where the gradients of $A$ and $\varphi$ vanish. Any change in $A$ will produce a compensating change in the charged particle momentum $Mv$ to preserve the conservation of generalized momentum along that path. Thus, $A$ is the momentum "stored" in the system comprised of a of a unit test charge in an external magnetic field.

One can now experimentally measure $A$ at all points in space by applying a series of concentric rings around the solenoid, each with a unit charge on a sliding bead. The vector potential is then obtained by monitoring the associated changes in the bead's momentum $Mv$ arising from solenoid's current changes and by applying conservation of momentum. Thus, the vector potential satisfies the criteria for a "real" field in the physical sciences: the vector potential $A$ has physical meaning and its characteristics are experimentally accessible at every point in space. One of Konopinski's major accomplishments is that both classical and quantum mechanics are now on the same footing regarding the primacy of the potential formulations. This difference in views of the reality of $A$ in classical and quantum physics was a major inconsistency that eventually had to be eliminated.

We now offer a simpler approach that demonstrates the essential points made by Konopinski without explicitly discussing the generalized momentum. Following convention, Eq.(3) is a law of physics in the standard gauge. One can now experimentally determine $A$ with the same solenoid setup by slowly increasing the solenoid's current from zero to its value at time t while concurrently monitoring the change in the bead's momentum. There are no external Coulomb fields, and no gradients in $A$ along the ring, so the force on the unit charged bead of mass $M$ is given by

$$F = d(Mv)/dt = -\partial A/\partial t = -dA/dt, \qquad (7)$$

giving $A(t) = -Mv$. In this simpler approach, $A$ is given physical meaning through Faraday's law of induction, and its measurability is demonstrated by using Konopinski's solenoid apparatus. Furthermore, this operational definition establishes its divergence: since $A = 0$ at t=0, and $\nabla \cdot E_I = 0 = \partial(\nabla \cdot A)/\partial t = 0$ throughout the measurement process, it follows that $\nabla \cdot A = 0$. This is physics, not the Coulomb gauge.

The next item relates to the inconsistency which Konopinski overlooked.  Elsewhere in his paper, he continues to maintain the view that one is free to choose the gauge for $A$. For example, in deriving the analytic expression for $A$ outside the infinite solenoid, he invokes the "Coulomb gauge" $\nabla \cdot A = 0$ in the usual manner implying that the divergence of $A$ is considered a "convenient choice".  So, on the one hand, he demonstrates that $A$ is physically real and measureable at all points in space, while on the other, $\nabla \cdot A$ is completely arbitrary.  This is a clear error since defining $A$ at all points in space also defines $\nabla \cdot A$.  (As we discussed above, his is essentially the same type of error that is discussed in reference [4].)  This contradiction is again resolved by recognizing that $\nabla \varphi_I$ =0 is an implicit gauge assumption in his analysis and one must avoid simultaneously invoke two potentially conflicting gauges.

We close with several comments.  Konopinski's demonstration that, by using the conventional formulation of Eq.(3), the vector potential possesses both physical significance and measurability shows the fallacy of the popular view that the vector potential is indefinable and is only a mathematical convenience with no physical significance.  The potentials now have the same primacy as they have in quantum mechanics because the vector potential is real and definable, even in regions where B=0.

The focus here is on situations where no dynamic Coulomb fields exist, so that $\nabla \cdot A = 0$.  Reference [3] extends the discussion of the consequence of the standard gauge assumption to more general cases and shows that

$$\nabla \bullet A = -(\partial \varphi / \partial t) / c^2 \tag{8}$$

is also a general law of physics in the standard gauge. It reflects a key physical principle at the core of electromagnetism.  It is not an arbitrary "condition".

The result that all variables are now precisely defined and measurable is also in better accord with the need for physical scientists to deal with matters that have physical meaning and experimental accessibility.  This approach also eliminates the source of fallacies and errors inherent in the current practice of applying gauge concepts; we include here the confusion that entrapped Jackson [4], and, ironically, Konopinski himself, as discussed above.  Another advantage of this approach is that it rids one of the need to entertain the broad variety of gauges (e.g., Lorenz gauge, Coulomb gauge, radiation gauge and local gauge).  Although the gauge concept plays a role in electromagnetism, the present analysis indicates it should be treated as a footnote, not a central feature.